\documentclass[preprint,prd,aps,tighten,nofootinbib,amssymb]{revtex4}
\usepackage{graphicx}
\usepackage{epsfig,amssymb,amsmath}

\newcommand{\beq}{\begin{equation}}
\newcommand{\eeq}{\end{equation}}
\newcommand{\bea}{\begin{eqnarray}}
\newcommand{\eea}{\end{eqnarray}}
\newcommand{\bear}{\begin{array}}
\newcommand {\eear}{\end{array}}
\newcommand{\bef}{\begin{figure}}
\newcommand {\eef}{\end{figure}}
\newcommand{\bec}{\begin{center}}
\newcommand {\eec}{\end{center}}

\begin{document}
\draft
\tighten
\preprint{CTPU-15-16}
\title{\large \bf Realizing the relaxion from multiple axions \\
and its UV completion with high scale supersymmetry}
\author{
    Kiwoon Choi\footnote{email: kchoi@ibs.re.kr} and
    Sang Hui Im\footnote{email: shim@ibs.re.kr}}
\affiliation{
 Center for Theoretical Physics of the Universe \\
 Institute for Basic Science (IBS), Daejeon 34051, Korea 
    }

\vspace{2cm}

\begin{abstract}

We discuss a scheme to implement the relaxion solution to the hierarchy problem with multiple axions, and present a UV-completed model realizing  the scheme. 
All of the $N$ axions in our model are periodic 
with a similar decay constant $f$ well below the Planck scale.  In the limit $N\gg 1$, the relaxion $\phi$ corresponds to an exponentially long multi-helical flat direction 
which is shaped by a  series of mass mixing between nearby axions
in the compact field space of $N$ axions. 
With the length of flat direction given by $\Delta \phi =2\pi f_{\rm eff} \sim e^{\xi N} f$ for $\xi={\cal O}(1)$,
both the scalar potential driving the evolution of $\phi$ during the inflationary epoch and the $\phi$-dependent Higgs boson mass vary  with an exponentially large periodicity of ${\cal O}(f_{\rm eff})$, while the back reaction potential stabilizing the relaxion has a periodicity of  ${\cal O}(
f)$. A natural UV completion of our scheme can be found in high scale or (mini) split supersymmetry (SUSY)  scenario  with the axion scales
generated by SUSY breaking as $f\sim \sqrt{m_{\rm SUSY}M_*}$, 
where  the soft SUSY breaking scalar mass $m_{\rm SUSY}$ can be well above the weak scale, and the fundamental scale $M_*$ can be identified as the Planck scale or the GUT scale.

\end{abstract}

\maketitle

\section{Introduction}

Recently a new approach to address the hierarchy problem has been proposed in \cite{relaxion}. The scheme introduces a  scalar degree of freedom, the relaxion
$\phi$, making
the Higgs boson mass a dynamical field depending on $\phi$. During the inflationary epoch, the Higgs boson mass-square $\mu_h^2(\phi)$ is scanned by the rolling $\phi$ from a large positive initial value  to zero. 
Right after the relaxion crosses the point $\mu_h(\phi)=0$, so that $\mu_h^2(\phi)$ becomes negative, a nonzero Higgs vacuum expectation value (VEV)  is developed and  a Higgs-dependent 
back reaction potential begins to operate to stabilize the relaxion\footnote{A mechanism to cosmologically relax the Higgs boson mass down to a small value through a nucleation of domain wall bubbles  has been discussed in \cite{dvali}.}.  One can then arrange the model parameters in a technically natural way to result in the relaxion stabilized at a point where the corresponding 
Higgs VEV is much smaller than the initial Higgs boson mass. 

An intriguing  feature of the relaxion mechanism is that the relaxion potential involves two very different scales. One is the period of the back reaction potential, and the other is the excursion range of the relaxion necessary to scan $\mu_h(\phi)$ from a large initial value to zero.
To see this, let us consider 
the relaxion potential given by
\bea
V(\phi, h) = V_0(\phi) + \mu_h^2(\phi) |h|^2 + V_{\rm br}(\phi, h)
\eea
where
$V_0$ is the potential driving the rolling of $\phi$ during the inflationary epoch
and  $V_{\rm br}$ is the periodic back reaction potential stabilizing $\phi$ right after it crosses $\mu_h(\phi)=0$. 
In fact, the key feature of the mechanism can be read off from the following  form of potential:
\bea
V_0 \,=\, \epsilon_0 f^3 \phi + ...., \quad
 \mu_h^2 \,=\, M_h^2 + \epsilon_h f \phi + ..., \quad V_{\rm br} \,=\, \Lambda^4_{\rm br}(h) \cos\left(\frac{\phi}{f}\right),\eea
 where $M_h$ denotes the initial Higgs boson mass, $\epsilon_0$ and $\epsilon_h$ are small dimensionless parameters 
 describing the explicit breaking of the relaxion shift symmetry in $V_0$ and $\mu^2_h$, respectively,  and finally $f$ is the relaxion decay constant in the back reaction potential.
 In non-supersymmetric  theory, the Higgs mass parameter $M_h$  is naturally of the order of the cutoff scale of the model.
 On the other hand, in supersymmetric theory, it corresponds to the scale of soft supersymmetry (SUSY) breaking mass which can be well below the cutoff scale of the model. In any case, we are interested in the case that $M_h$ is much larger than the weak scale:
 \bea
M_h \,\gg \,    v\equiv \langle h\rangle =174 \,\, {\rm GeV},\eea
which might be explained by the relaxion mechanism. 

Let us now list the conditions for the relaxion mechanism to work. First of all,
in order for the rolling relaxion to cross $\mu_h(\phi)=0$ without a fine tuning of the initial condition, 
it should experience a  field excursion    
\bea
\frac{\Delta \phi}{f} \, \gtrsim  \, \frac{M^2_h}{\epsilon_h f^2}.\eea
In order for the scalar potential to be technically natural under  radiative corrections, the symmetry breaking parameters $\epsilon_0$ and $\epsilon_h$ should obey
\bea
\epsilon_0 \, \gtrsim\, \epsilon_h  \frac{M_h^2}{f^2}.
\eea
On the other hand, from the stability condition $\partial_\phi V=0$, one finds \bea
\epsilon_0 \,\sim\, \frac{\Lambda^4_{\rm br}}{f^4},\eea
and therefore
\bea \label{excur}
\frac{\Delta \phi}{f} \, \gtrsim\, \frac{M_h^4}{\Lambda_{\rm br}^4}.\eea
As for the back reaction potential, generically $\Lambda_{\rm br}(h=0)$ may not be vanishing, and then
one needs
\bea
\Lambda^4_{\rm br}(h=v) \,\gg\, \Lambda^4_{\rm br}(h=0).\eea
Also, in order not to destabilize the weak scale size of the Higgs VEV, its magnitude should be bounded as
\bea
\label{br_bound}
\Lambda_{\rm br}(h=v) \, \lesssim \, {\cal O}(v).\eea

An immediate consequence of the above conditions is that
the relaxion should experience a field excursion much bigger than $f$ in the limit $M_h\gg v$:
\bea
\frac{\Delta \phi}{f}\, \gtrsim\, \frac{M_h^4}{v^4}.\eea
The required excursion  is huge in the case that the  back reaction potential is generated by the QCD anomaly, in which
$\Lambda^4_{\rm br}\,\sim\, f_\pi^2 m_\pi^2$ and therefore
\bea
\frac{\Delta \phi}{f} \, \gtrsim\, {\cal O}\left(\frac{M_h^4}{f^2_\pi m_\pi^2}\right)\,\sim\,  10^{12}\left(\frac{M_h}{v}\right)^4.\eea
Even when the scale of the back reaction potential saturates the bound (\ref{br_bound}), the required relaxion excursion
is still much larger than $f$  as long as $M_h$ is higher than the weak scale 
by more than a few  orders of magnitudes.
Note that the natural size of  $M_h$ 
is the cutoff scale of the model for non-SUSY case, while it is the soft SUSY breaking scalar mass for SUSY case. 

Therefore, in the relaxion scenario, the hierarchy $M_h/v \gg 1$  is replaced with a much bigger hierarchy $\Delta \phi/f \gtrsim M^4_h/v^4$. Although $\Delta \phi \gg f$ might be stable against radiative corrections, it is still crying  for an explanation with a sensible UV completion.
To incorporate a huge relaxion excursion, one may simply assume that the relaxion is a non-compact field variable.
See \cite{Espinosa:2015eda, Hardy:2015laa, Patil:2015oxa, Jaeckel:2015txa, Gupta:2015uea, Batell:2015fma, Matsedonskyi:2015xta, Marzola:2015dia} for  recent discussions  of the related issues.  In this paper, we  discuss an alternative scenario in which the relaxion corresponds to an exponentially long multi-helical flat direction 
in the compact field space spanned by $N$ sub-Planckian periodic axions:
$$ 
\phi_i\equiv \phi_i+2\pi f_i \quad (i=1,2,..,N) 
$$
with  $f_i\ll M_{\rm Planck}$. Such a long flat direction 
 is formed 
 by a series of mass mixing between nearby axions, producing a   multiplicative sequence of helical windings of flat direction, which results in 
$$
\frac{\Delta \phi}{f_i} \,=\, {\cal O}(e^{\xi N}) $$
for $\xi={\cal O}(1)$.
 Our scenario is inspired by the recent generalization of the axion alignment mechanism for natural inflation
 \cite{Kim-Nilles-Peloso} to the case of $N$ axions
\cite{Choi-Kim-Yun}.
Although it requires a rather specific form of  axion mass mixings, our scheme 
 does not  involve any fine tuning of continuous parameters, nor an unreasonably large discrete parameter.


As we will see, our scheme finds a natural UV completion in high scale or (mini) split supersymmetry (SUSY) scenario with soft SUSY breaking scalar mass $m_{\rm SUSY}\gg v$.  In the UV completed model, the axion scales are generated by SUSY breaking \cite{Kim-Nilles, Murayama-Suzuki-Yanagida, Choi-Chun-Kim}
as 
$$f_i \sim \sqrt{m_{\rm SUSY}M_*},$$
where $M_*$ can be identified as the Planck scale or the GUT scale.
With the $(N-1)$ hidden Yang-Mills gauge sectors which confine at scales below $f_i$ to generate the desired axion mass mixings, the
 canonically normalized relaxion  has a field range 
 $$\Delta \phi \equiv 2\pi f_{\rm eff} \sim  2\pi f_i\left(\prod_{j=1}^{N-1} n_j\right),$$
  where
$n_j>1$  corresponds to the number of flavors of the gauge-charged fermions in the $j$-th hidden sector.  
One can then arrange the microscopic parameters in a technically natural way  to make 
the resulting relaxion potential $V_0(\phi)$ and the $\phi$-dependent Higgs boson mass $\mu_h^2(\phi)$  vary with an exponentially large  periodicity of  ${\cal O}(f_{\rm eff})$, while the back reaction potential $V_{\rm br}(h,\phi)$ has a periodicity of ${\cal O}(f_i)$. An interesting feature of our model is that the desired  $V_0(\phi)$ and $\mu_h^2(\phi)$ arise as a natural consequence of the solution of the MSSM $\mu$-problem advocated in \cite{Kim-Nilles, Murayama-Suzuki-Yanagida, Choi-Chun-Kim}.
 
The outline of the paper is as follows.
In the next section, we describe the basic idea with a simple toy model and discuss the scheme  within the framework of an effective theory of $N$ axions. In section 3, we present a UV model with high scale SUSY, realizing our scheme in the low energy limit. Section 4 is the conclusion.

\section{exponentially long relaxion from multiple axions}

To illustrate the basic idea, let us begin with a simple two axion model.
The lagrangian density of the model is given by 
\bea
{\cal L}\,=\, \frac{1}{2}\left(\partial_\mu \phi_1\right)^2 +\frac{1}{2}\left(\partial_\mu \phi_2\right)^2
-\left( \tilde V_0 +V_0+ \mu_h^2|h|^2 + V_{\rm br}+...\right),\eea
where  $h$ is the
 SM Higgs doublet and $\phi_i$ are the periodic axions:
\bea
\phi_i \,\equiv\, \phi_i +2\pi f_i,\eea
with a scalar potential 
\bea
\tilde V_0&=& -\Lambda^4\cos \left(\frac{\phi_1}{f_1}+n \frac{\phi_2}{f_2}\right),\nonumber \\
V_{0}&=&
-\epsilon f_2^4 \cos\left(\frac{\phi_2}{f_2}+\delta_2\right),
\nonumber
 \\
 \mu_h^2&=& M_h^2-\epsilon^\prime f_2^2 \cos\left(\frac{\phi_2}{f_2}+\delta_2^\prime\right),\nonumber \\
 V_{\rm br}&=& -\Lambda^4_{\rm br}(h)\cos\left(\frac{\phi_1}{f_1}+\delta_1\right),
\eea
where \bea
\Lambda^4 \,\gg\, \epsilon f_2^4 \,\gg\, \Lambda_{\rm br}^4.\eea
Here $M_h$ is an {\it axion-independent} mass parameter which  is comparable to the cutoff scale of the above effective lagrangian, and $n>1$ is an integer which will be determined by the underlying UV completion.
We assume
\bea \label{cond_1}
 \epsilon f_2^2 \, \gtrsim\, {\cal O}(\epsilon^\prime M_h^2), \,\quad 
\epsilon^\prime f_2^2 \, \gtrsim\, {\cal O}(M_h^2), 
\eea
and therefore the model  is stable against the radiative corrections which replace the
Higgs operator $|h|^2$ with the cutoff-square  of ${\cal O}(M_h^2)$, while allowing $\mu_h^2=0$ for certain value of $\phi_2$. 

As for the back reaction potential, one can consider two different possibilities. One option is to generate it by the coupling of $\phi_1$ to the QCD anomaly, yielding
\bea
\Lambda_{\rm br}^4(h) \,\sim\, y_u \Lambda_{\rm QCD}^3 h,\eea
where $y_u$ denotes the up quark Yukawa coupling to the SM Higgs field $h$, and $\Lambda_{\rm QCD}$ is the QCD scale. 
This option corresponds to the minimal model, however generically is in conflict with the axion solution to the strong CP problem. Alternative option is to introduce a new hidden gauge interaction which confines around the weak scale and generates
a back reaction potential given by \cite{relaxion, Antipin:2015jia}
\bea
\Lambda_{\rm br}^4 \, =\, m_1^2 |h|^2 + m_2^4 \eea
with
\bea
m_2^4 \,<\, m_1^2 v^2  \, \lesssim\, {\cal O}(v^4).
\eea
In order for the model to be technically natural, the underlying dynamics to generate the back reaction potential should be arranged to
make sure that the above conditions on $m_1$ and $m_2$  are stable against radiative corrections.

\begin{figure}[t]
\begin{center}
 \begin{tabular}{l}
  \includegraphics[width=0.9 \textwidth]{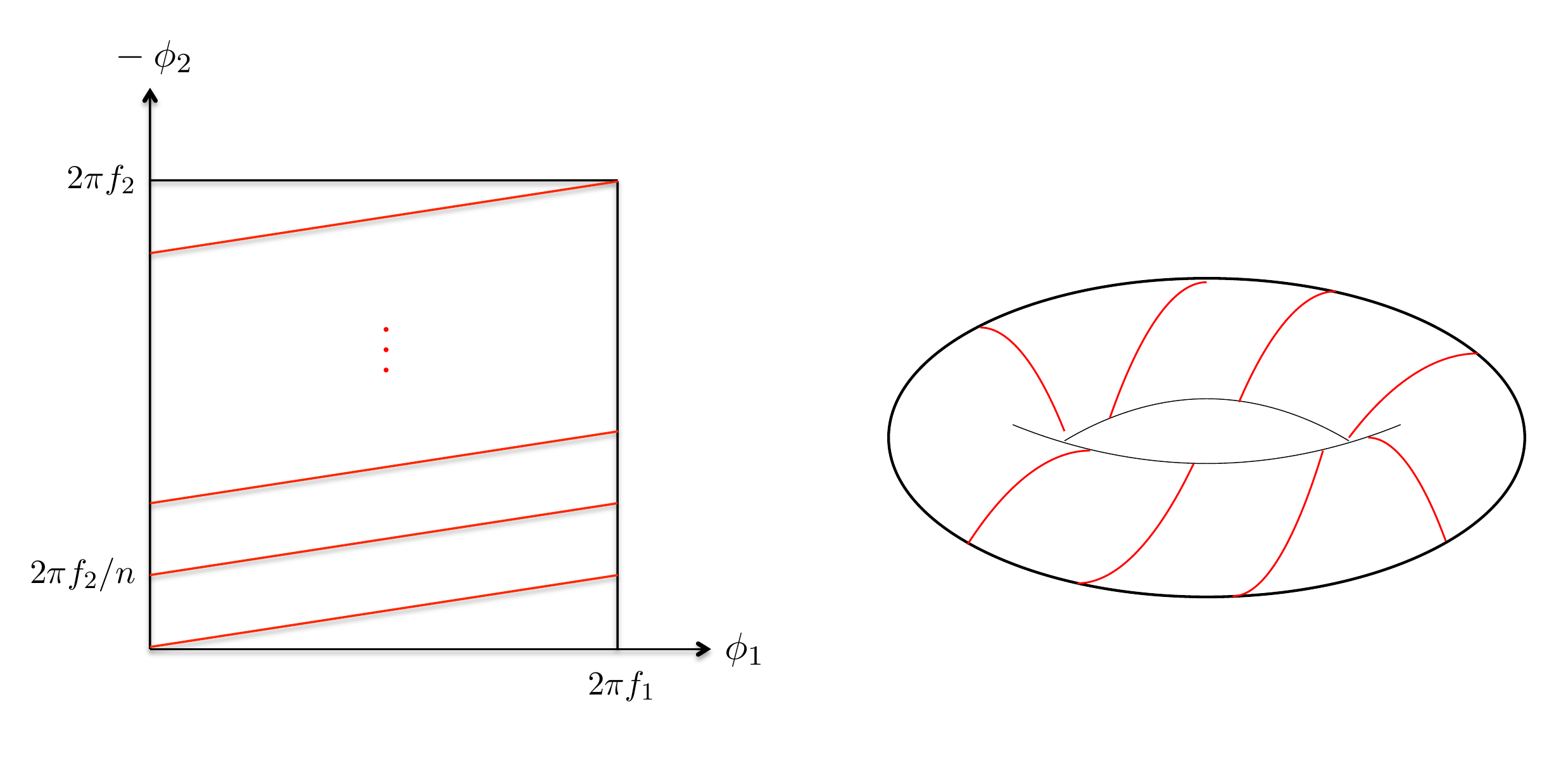}
   \end{tabular}
  \end{center}
  \caption{
Flat relaxion direction in the two axion model.
  }
\label{fig:helical}
\end{figure}

The above two axion model involves the shift symmetries 
\bea
U(1)_i: \quad \frac{\phi_i}{f_i}\,\rightarrow \frac{\phi_i}{f_i} + c_i \quad (i=1,2),
\eea
which are broken by $\tilde V_0$ down to the relaxion shift symmetry
\bea
U(1)_\phi:\quad \frac{\phi_1}{f_1} \rightarrow \frac{\phi_1}{f_1} +nc, \quad 
\frac{\phi_2}{f_2} \rightarrow \frac{\phi_2}{f_2}- c.
\eea
The flat direction associated with $U(1)_\phi$ 
has a helical winding structure in the compact 2-dim field space of $\phi_i$ as depicted in Fig. (\ref{fig:helical}). 
Then the periodicity of the flat direction is enlarged as
\bea
\Delta \phi \,=\, 2\pi \sqrt{n^2 f_1^2 +f_2^2}\,\equiv 2\pi f_{\rm eff},
\eea
which is larger than the original axion periodicities $2\pi f_1 \sim\, 2\pi f_2$ by the winding number $n$.

The relaxion shift symmetry $U(1)_\phi$ is  slightly broken by small nonzero values of $\epsilon, \epsilon^\prime$ and $\Lambda_{\rm br}$. Note that this particular form of $U(1)_\phi$ breaking is technically natural as long as the first condition
of (\ref{cond_1}) is satisfied.  
To find the effective potential of the flat relaxion direction, one can rewrite the model 
in terms of the canonically normalized heavy and light axions \cite{Kim-Nilles-Peloso,Choi-Kim-Yun}:
\bea
\phi_H \,=\, \frac{f_2\phi_1 +nf_1 \phi_2}{f_{\rm eff}}, \quad
\phi \,=\, \frac{nf_1\phi_1-f_2 \phi_2}{f_{\rm eff}},\eea
for which
\bea
\frac{\phi_1}{f_1} &=& n\frac{\phi}{f_{\rm eff}}+\frac{f_2^2}{n^2f_1^2+f_2^2} \frac{\phi_H}{f_H}\nonumber \\
\frac{\phi_2}{f_2} &=& -\frac{\phi}{f_{\rm eff}}+\frac{nf_1^2}{n^2f_1^2+f_2^2} \frac{\phi_H}{f_H},\eea
where $f_H=f_1f_2/f_{\rm eff}$.
In the limit $\Lambda^4 \gg \epsilon f_2^4 \gg  \Lambda^4_{\rm br}$, it is straightforward to integrate out the heavy axion $\phi_H$ to derive the low energy effective lagrangian of the light axion $\phi$. The resulting effective potential
of the canonically normalized $\phi$ is given by 
\bea
V_{\rm eff} &=&
 -\epsilon f_2^4\cos\left(\frac{\phi}{f_{\rm eff}}-\delta_2\right)
 +\left(M_h^2 -\epsilon^\prime f_2^2 \cos\left(\frac{\phi}{f_{\rm eff}}-\delta_2^\prime\right)\right)|h|^2\nonumber \\
 &&-\, \Lambda_{\rm br}^4(h) \cos\left(\frac{\phi}{f}+\delta_1\right), \eea
where
\bea
f_{\rm eff} \,=\, \sqrt{n^2f_1^2+f_2^2}\,\equiv nf.\eea

We can now generalize the above two axion model 
to the case of $N>2$ axions to enlarge the effective axion scale further \cite{Choi-Kim-Yun}. The lagrangian density is given by
\bea
{\cal L}\,=\, \frac{1}{2}\sum_i \left(\partial_\mu\phi_i\right)^2 -\left( \tilde V_0 +V_0+ \mu_h^2|h|^2 + V_{\rm br}+...\right),
\eea
where
\bea \label{potentials}
\tilde V_0 &=& -\sum_{i=1}^{N-1}\Lambda_i^4 
\cos \left(\frac{\phi_i}{f_i}+n_{i} \frac{\phi_{i+1}}{f_{i+1}}\right)\nonumber \\
V_{0}&=&
-\epsilon f_N^4 \cos\left(\frac{\phi_N}{f_N}+\delta_N\right),
\nonumber
 \\
 \mu_h^2&=& M_h^2-\epsilon^\prime f_N^2 \cos\left(\frac{\phi_N}{f_N}+\delta_N^\prime\right),\nonumber \\
 V_{\rm br}&=& -\Lambda^4_{\rm br}(h)\cos\left(\frac{\phi_1}{f_1}+\delta_1\right), 
\eea
with $\Lambda_i^4  \gg \epsilon f_N^4 \gg \Lambda_{\rm br}^4$. The model involves the $N$ axionic shift symmetries:
\bea
U(1)_i:\quad 
\frac{\phi_i}{f_i}\, \rightarrow \, \frac{\phi_i}{f_i} + c_i\eea
which are broken by $\tilde V_0$ down to the relaxion shift symmetry:
\bea
\label{relaxion_shift}
U(1)_\phi:\quad \frac{\phi_i}{f_i}\, \rightarrow \, \frac{\phi_i}{f_i} +\gamma_i c \quad (\gamma_i=-n_i\gamma_{i+1}),\eea
with the corresponding flat direction given by 
\bea
\phi \, \propto \, \sum_{i=1}^{N} (-1)^{i-1}\left(\prod_{j={i}}^{N-1} n_j\right) f_i \phi_i.\eea

Turing on small nonzero values of $\epsilon, \epsilon^\prime$ and $\Lambda_{\rm br}$, the relaxion shift symmetry (\ref{relaxion_shift}) is slightly broken and nontrivial potential of $\phi$ is developed.  Although our way to break $U(1)_\phi$ is rather specific, it is technically natural as the model involves many continuous or discrete axionic 
shift symmetries which are distinguishing our particular way of symmetry  breaking from other possibilities.  
It is again straightforward to integrate out the $(N-1)$ heavy axions which receive a large mass from $\tilde V_0$ \cite{Choi-Kim-Yun}.
For the canonically normalized $\phi$, the resulting effective potential  is given by
\bea \label{Veff}
V_{\rm eff} &=& V_0(\phi) +\mu_h^2(\phi)|h|^2 +V_{\rm br}(h,\phi)\nonumber \\
&=& -\epsilon f_N^4\cos\left(\frac{\phi}{f_{\rm eff}}+(-)^{N-1}\delta_N\right)
 +\left(M_h^2 -\epsilon^\prime f_N^2 \cos\left(\frac{\phi}{f_{\rm eff}}+(-)^{N-1}\delta_N^\prime\right)\right)|h|^2
 \nonumber \\
&&-\, \Lambda_{\rm br}^4(h) \cos\left(\frac{\phi}{f}+\delta_1\right), \eea
where
\bea \label{feff}
f_{\rm eff} &=& \sqrt{\sum_{i=1}^N \left(\prod_{j=i}^{N-1} n_j^2\right) f_i^2}\,\sim\, \left(\prod_{j=1}^{N-1} n_j\right) f_1\,\sim\, e^{\xi N} f_1 \quad \left(\xi={\cal O}(1)\right),\nonumber \\
f&=& \frac{f_{\rm eff}}{\left(\prod_{j=1}^{N-1} n_j\right)}\,\sim\, f_1. \eea
For simplicity here we assumed that all $f_i$ are comparable to each other, or $f_1$ is the biggest among $\{f_i\}$.

\begin{figure}[t]
  \begin{center}
   \begin{tabular}{l}
   \includegraphics[width=0.7 \textwidth]{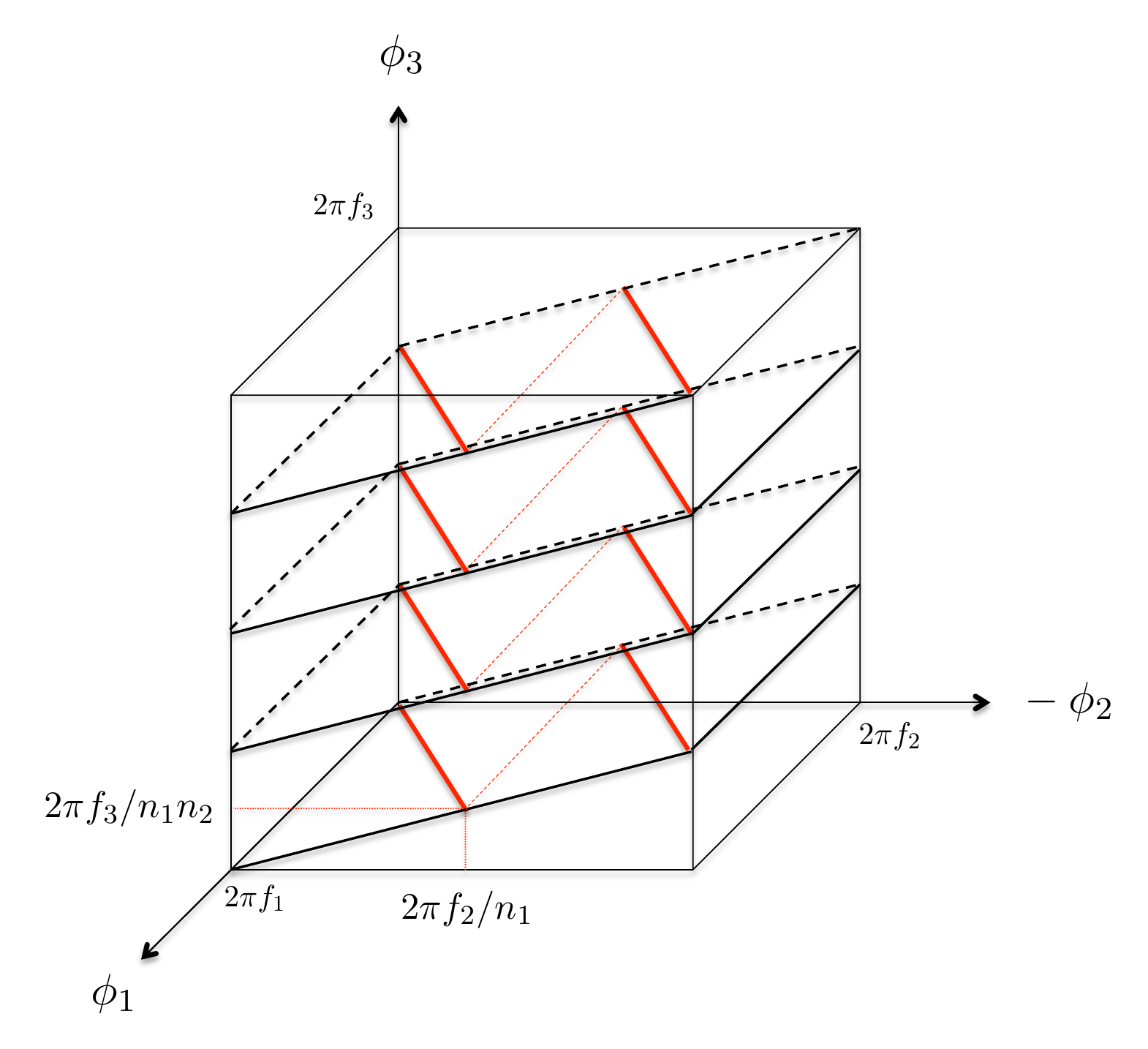}
   \end{tabular}
  \end{center}
  \caption{
   Flat relaxion direction in the three axion case with $n_1=2$ and $n_2=4$.
   }
 \label{fig:multi}
 \end{figure}

Obviously,  in the limit $N\gg 1$ the above relaxion potential involves two very different axion scales, an exponentially enhanced effective decay constant $f_{\rm eff}$  and another effective decay constant $f$ which
is comparable to the original  decay constants $f_i$. Such a big difference between $f_{\rm eff}$ and $f$ 
can be understood by noting  that in order for the $N$-th axion $\phi_N$  to travel one period
along the relaxion direction, i.e. $\Delta \phi_N = 2\pi f_N$, the other axions $\phi_i$ $(i=1,2,..., N-1)$ should experience a multiple windings as $\Delta \phi_i = 2\pi \left(\prod_{j=i}^{N-1} n_j\right) f_i$.
This results in 
\bea
\frac{\phi_i}{f_i} \, =\, (-1)^{i-1} \left(\prod_{j=i}^{N-1} n_j\right)\frac{\phi}{f_{\rm eff}} + ...,\eea
where the ellipsis stands for the $(N-1)$ heavy axions  receiving a large mass from $\tilde V_0$.
For an illustration of this feature, we depict in Fig. (\ref{fig:multi}) the relaxion field direction  for the case of $N=3, n_1=2, n_2=4$.

The effective potential (\ref{Veff}) can easily realize the relaxion mechanism under several consistency conditions. 
First of all, like (\ref{cond_1}) of the two axion model, we need
\bea \label{cond_1'}
 \epsilon f_N^2 \, \gtrsim\, {\cal O}(\epsilon^\prime M_h^2), \,\quad 
\epsilon^\prime f_N^2 \, \gtrsim\, {\cal O}(M_h^2), 
\eea
in order for the model to be stable against radiative corrections, while allowing $\mu_h = 0$ for certain value of $\phi$.
Without invoking any fine tuning, there is always a certain range of  $\delta_N$ and $\delta^\prime_N$ for which the relaxion rolls down toward the minimum of $V_0(\phi)$ starting 
from a generic initial value $\phi_0$ with $\mu_h^2(\phi_0) ={\cal O}(M_h^2) > 0$.
After a field excursion $\Delta \phi={\cal O}(f_{\rm eff})$, the relaxion is crossing  $\mu_h(\phi)=0$, and
then a nonzero Higgs VEV is developed together with the back reaction potential stabilizing the relaxion at the value giving $\langle h\rangle = v$.
The stabilization condition leads to
\bea
\frac{\epsilon f_N^4}{f_{\rm eff}} \,\sim\,  \frac{\Lambda_{\rm br}^4(h=v)}{f}. 
\eea  
From (\ref{cond_1'}), this then yields a lower bound on $f_{\rm eff}$:
\bea \label{excur'}
\frac{f_{\rm eff}}{f} \, \gtrsim \, \frac{M_h^4}{\Lambda_{\rm br}^4(h=v)}\,=\,  \left(\frac{M_h}{v}\right)^4 \frac{v^4}{\Lambda_{\rm br}^4(h=v)},
\eea
where $v^4/\Lambda^4_{\rm br}(h=v)\sim 10^{12}$ when $V_{\rm br}$ is generated by the QCD anomaly, or 
$v^4/\Lambda^4_{\rm br}(h=v)$ has a model-dependent value not exceeding  ${\cal O}(1)$ when $V_{\rm br}$ is generated by the hidden color dynamics which confines around the weak scale.

To summarize,
in our scheme for the relaxion mechanism, $v\ll M_h$  can be technically natural with an exponential hierarchy between 
the two effective axion scales:
\bea
\frac{f_{\rm eff}}{f} \,=\, {\cal O}(e^{\xi N}) \quad (\xi={\cal O}(1))\eea
which is arising as a consequence of a series of mass mixing between nearby axions in the compact field space of $N$ axions.
Although it relies on a rather specific form of axion mass mixings,
the scheme  does not involve any fine tuning of continuous parameters, nor an unreasonably large discrete parameter.


\section{A UV model with high scale supersymmetry}

In this section, we construct an explicit UV completion of the $N$ axion model discussed in the previous section.  
We first note that our scheme requires that the axion potential $\tilde V_0$  should dominate over the other part
of the potential in (\ref{potentials}) as it determines the key feature of the model, i.e. an exponentially long flat direction in the compact field space of $N$ axions.  Specifically we need 
\bea
f_i^4 \,\gg\, |\tilde{V}_0|\,\gg\, |V_0|\, \gtrsim M_h^4.
\eea
On the other hand, we wish to have an explicit UV model providing the full part of the axion potential in (\ref{potentials}), as well as a mechanism to generate the axion scales $f_i$. This implies that our UV model should allow the natural size of the Higgs boson mass, i.e. $M_h$, to be much lower than its cutoff scale.
As SUSY provides a natural framework for this purpose,
in the following we present a supersymmetric UV completion of the low energy effective potential (\ref{Veff}).

First of all, to have $N$ axions with the decay constants $f_i\ll M_{\rm Planck}$, we introduce $N$ global $U(1)$ symmetries
under which 
\bea
U(1)_i:\quad X_i\rightarrow e^{i\beta_i}X_i, \quad Y_i\rightarrow e^{-3i\beta_i}Y_i \quad (i=1,2,...,N),\eea
where $X_i$ and $Y_i$ are gauge-singlet chiral superfields with the $U(1)_i$-invariant superpotential
\bea
W_1 \,=\, \sum_i \frac{X_i^3Y_i}{M_*},\eea
where $M_*$ corresponds to the cutoff scale of the model, which might be identified as  the GUT scale or the Planck scale.
Here and in the following, 
we ignore the dimensionless coefficients of order unity in the lagrangian.
We assume that SUSY is softly broken with SUSY breaking soft masses
\bea 
m_{\rm SUSY}\,\sim\, M_h\,\ll \, M_*.\eea
In particular, the model involves the soft SUSY breaking terms of $X_i$ and $Y_i$, given by
\bea
{\cal L}_{\rm soft}\,=\, -m_{X_i}^2|X_i|^2 - m_{Y_i}^2|Y_i|^2 +\left(A_i\frac{X_i^3Y_i}{M_*}+{\rm h.c}\right),\eea
where $$
m_{X_i}\,\sim\, m_{Y_i}\,\sim \, A_i\,\sim \, m_{\rm SUSY}.
$$ 
To achieve the $N$ axions in the low energy limit, we need all $m_{X_i}^2$ are tachyonic, which  results in
\bea
\langle X_i \rangle \,\equiv x_i \,\sim\, \sqrt{m_{\rm SUSY}M_*}, \quad \langle Y_i\rangle \,\equiv y_i\, \sim \sqrt{m_{\rm SUSY}M_*}.
\eea
Then the canonically normalized axion components $\phi_i$ can be identified as
\bea
X_i \,\propto\, e^{i\phi_i/f_i}, \quad Y_i\,\propto\, e^{-3i\phi_i/f_i}\eea
with
\bea
f_i \,=\, \sqrt{2(x_i^2+9y_i^2)}\,\sim\, \sqrt{m_{\rm SUSY}M_*}.\eea

Now we need a dynamics to generate the axion potential $\tilde V_0$ in (\ref{potentials}), developing an exponentially long  flat direction as described in the previous section. For this purpose, we introduce $(N-1)$ hidden Yang-Mills sectors associated with the gauge group $G=\prod_{i=1}^{N-1}SU(k_i)$, including also the charged matter fields
\bea
\Psi_i+\Psi_i^c, \quad \Phi_{ia}+\Phi_{ia}^c \quad (i=1,2,..., N-1; \, a=1,2,..., n_i),\eea
where $\Psi_i$ and $\Phi_{ia}$ are the fundamental representation of $SU(k_i)$, while $\Psi_i^c$ and $\Phi_{ia}^c$ are anti-fundamentals. These gauged charged matter fields
couple to the $U(1)_i$-breaking  fields $X_i$ through the superpotential
\bea
W_2 \,=\, \sum_{i=1}^{N-1}\left( X_i \Psi_i\Psi_i^c + X_{i+1}\Phi_{ia}\Phi^c_{ia}\right). 
\eea 
Note that $X_i$ couples to a single flavor of the $SU(k_i)$-charged hidden quark $\Psi_i+\Psi_i^c$, while $X_{i+1}$ couples to 
 $n_i$ flavors of the $SU(k_i)$-charged hidden quarks $\Phi_{ia}+\Phi^c_{ia}$.
With this form of hidden Yang-Mills sectors,  the $N$ global  $U(1)$ symmetries are explicitly broken down to 
a single $U(1)$ by the $U(1)_i\times SU(k_j)\times SU(k_j)$ anomalies. 
The charged matter fields $\Psi_i+\Psi_i^c$ and $\Phi_{ia}+\Phi_{ia}$ get a heavy mass of ${\cal O}(f_i)$, so can be integrated out at scales below $f_i$. This yields  an axion-dependent threshold correction to the holomorphic gauge kinetic function
$\tau_i$ of $SU(k_i)$
at scales below $f_i$:
\bea
\tau_i \,=\, \frac{1}{g_i^2}+\frac{i}{8\pi^2}\left(\frac{\phi_i}{f_i}+n_i\frac{\phi_{i+1}}{f_{i+1}}\right)+\theta^2M_{\lambda_i},\eea
where we ignored the dependence on $|X_i|$, while including the soft SUSY breaking by the gaugino masses $M_{\lambda_i}\sim m_{\rm SUSY}$.
As a consequence, at scales below $f_i$,
the global symmetry breaking by the $U(1)_i\times SU(k_j)\times SU(k_j)$ anomalies
is described by the following axion effective interactions:
\bea
\sum_{i=1}^{N-1}\frac{1}{32\pi^2}\left(\frac{\phi_i}{f_i}+n_i\frac{\phi_{i+1}}{f_{i+1}}\right)\left(F\tilde F\right)_{SU(k_i)},\eea
where $(F)_{SU(k_i)}$ denotes the gauge field strength of $SU(k_i)$ and $(\tilde{F})_{SU(k_i)}$ is its dual.
As we wish to generate the axion potential $|\tilde V_0|\gg M_h^4\sim m_{\rm SUSY}^4$ from the above axion couplings, 
we assume 
\bea
\tilde \Lambda_i \,\gg \, m_{\rm SUSY},\eea
where $\tilde\Lambda_i$ denotes the confining scale of the hidden gauge group $SU(k_i)$.
In such case, the resulting  axion potential is determined by the non-perturbative effective superpotential describing the formation of the $SU(k_i)$ gaugino condensation \cite{Intriligator:1995au}:
\bea
W_{\rm np}\,\sim\,  \langle \lambda_i\lambda_i\rangle \, \propto \left(e^{-8\pi^2 \tau_i}\right)^{1/k_i},
\eea
yielding
\bea
\tilde V_0 \,=\, -\sum_{i=1}^{N-1}\Lambda_i^4 \cos\left(\frac{1}{k_i}\left(\frac{\phi_i}{f_i}+n_i\frac{\phi_{i+1}}{f_{i+1}}\right)\right)\eea
with
\bea
\Lambda_i^4 \,\sim\, \frac{8\pi^2}{k_i} M_{\lambda_i}\tilde\Lambda_i^3.\eea

Our next mission is to generate the axion potential $V_0$ and the axion-dependent Higgs mass-square $\mu_h^2$ in (\ref{potentials}), 
driving the evolution of the relaxion during the inflationary epoch,  while scanning  the Higgs mass-square
from an initial value of ${\cal O}(m_{\rm SUSY}^2)$ to zero.
This can be done by introducing a superpotential term given by
\bea
W_3 &=& \left(\frac{X_{N-1}^2}{M_*} + \frac{X_N^2}{M_*} \right) H_u H_d, 
\eea 
together with the associated  K\"ahler potential term:
\bea
\Delta K \,=\,  \frac{X_{N-1}^2X_N^{*2}}{M_*^2}+{\rm h.c}.\eea
Here we ignore the irrelevant terms such as $|X_N|^4$ or $|X_{N-1}|^4$ in the K\"ahler potential. Note that 
the couplings in $W_3$ leads to a logarithmically divergent radiative correction to $\Delta K$ \cite{Grisaru:1996ve}, and our model is stable  against such radiative correction as long as the coefficient of $\Delta K$ is of order unity. Note also that $W_3$ provides a
solution to the MSSM $\mu$ problem as it yields naturally the Higgsino mass $\mu_{\rm eff}\,\sim\, m_{\rm SUSY}$ \cite{Kim-Nilles,
Murayama-Suzuki-Yanagida, Choi-Chun-Kim}.

After integrating out the $(N-1)$ axions which receive a heavy mass from $\tilde V_0$, while leaving the light relaxion $\phi$ as described in the previous section, 
the K\"ahler potential term $\Delta K$ gives rise to
\bea
V_0 \,=\, -m_0^4 \cos\left( 2(n_{N-1}+1)\frac{\phi}{f_{\rm eff}} +\delta \right),\eea
where
\bea
m_0^4 &\sim& \frac{f_{N-1}^2f_N^2}{M_*^2} m_{\rm SUSY}^2 \,\sim\, m_{\rm SUSY}^4,\nonumber \\
f_{\rm eff} &=& \sqrt{\sum_{i=1}^N \left(\prod_{j=i}^{N-1} n_j^2\right) f_i^2}\,\sim\, \left(\prod_{j=1}^{N-1} n_j\right) f_1, \eea
and $\delta$ is a phase angle which is generically of order unity.
In our scheme,  the MSSM Higgsino mass $\mu_{\rm eff}$ originates from $W_3$, and therefore is naturally of the order of $m_{\rm SUSY}$ \cite{Kim-Nilles, Murayama-Suzuki-Yanagida, Choi-Chun-Kim}.
Again, after integrating out the $(N-1)$ heavy axions, 
we find the MSSM Higgs parameters $\mu_{\rm eff}$ and $B\mu_{\rm eff}$ depend on the relaxion $\phi$ as
\bea \label{mu_b}
\mu_{\rm eff} &=& \mu_{N-1} \exp(- i 2 n_{N-1} \phi/f_{\rm eff})+ \mu_N \exp(i 2  \phi/f_{\rm eff}), \nonumber \\
B\mu_{\rm eff} &=&  b_{N-1} \exp(- i 2 n_{N-1} \phi/f_{\rm eff}) + b_N \exp(i 2 \phi/f_{\rm eff}),
\eea
where
\bea
|\mu_N|\, \sim\, |\mu_{N-1}| \,\sim\, \frac{f^2}{M_*}\,\sim\, m_{\rm SUSY}, \quad |b_N|\,\sim\, |b_{N-1}|\,\sim\, m_{\rm SUSY}^2.\eea
Then the determinant of the MSSM Higgs mass matrix
\bea
{\cal D} = (m_{H_u}^2 + |\mu_{\rm eff}|^2 ) (m_{H_d}^2 + |\mu_{\rm eff}|^2 ) - |B\mu_{\rm eff}|^2
\eea
also depends on $\phi$ via
\bea \label{mu_b^2}
|\mu_{\rm eff}|^2 &=& |\mu_N|^2 + |\mu_{N-1}|^2 +2|\mu_{N}\mu_{N-1}|\cos\left( 2(n_{N-1}+1)\frac{\phi}{f_{\rm eff}} + \delta_{\mu_N} - \delta_{\mu_{N-1}}\right), \nonumber\\
|B\mu_{\rm eff}|^2 &=& |b_N|^2 + |b_{N-1}|^2 +2|b_{N}b_{N-1}|\cos\left( 2(n_{N-1}+1)\frac{\phi}{f_{\rm eff}} + \delta_{b_N} - \delta_{b_{N-1}}\right), 
\eea
where $\delta_{\mu}$ and $\delta_{b}$ are the phases of $\mu$ and $b$, respectively. 
Obviously, for an appropriate range of $\delta_\mu$ and $\delta_b$,  the determinant ${\cal D}$
can flip its sign from positive to negative as the relaxion experiences an excursion of ${\cal O}(f_{\rm eff})$.
Once the relaxion is stabilized near
the point of ${\cal D}=0$, the MSSM Higgs doublets $H_u$ and $H_d$ can be decomposed into the light SM Higgs $h$ with a mass
of ${\cal O}(v)$
and the other heavy Higgs bosons having a mass of the order of $m_{\rm SUSY}\gg v$.

To complete the model, we need to generate the back reaction potential $V_{\rm br}$. In regard to this, we simply adopt  the schemes suggested in \cite{relaxion}. One option is
to generate $V_{\rm br}$ through the QCD anomaly. For this, one can  introduce
\bea
W_{\rm br} \,=\, X_1 QQ^c,\eea
where $Q+Q^c$ is an exotic quark which receive a heavy mass by $\langle X_1\rangle \sim f_1$.
Once this heavy quark is integrated out, the axion $\phi_1$ couples to the gluons as
\bea
\frac{1}{32\pi^2}\frac{\phi_1}{f_1}\left(F\tilde F\right)_{\rm QCD}.
\eea
After the  $(N-1)$ heavy axions are integrated out, this leads to the relaxion-gluon coupling
\bea
\frac{1}{32\pi^2}\frac{\phi}{f}\left(F \tilde F\right)_{\rm QCD},
\eea
where
\bea
f\,=\, \frac{f_{\rm eff}}{\left(\prod_{j=1}^{N-1} n_j\right)} \,\sim\, f_1.\eea
Then the resulting back reaction potential is obtained to be
\bea
V_{\rm br}(h,\phi)\,\approx\, -y_u\Lambda^3_{\rm QCD}h \cos\left(\frac{\phi}{f} +\delta_{\rm br}\right),
\eea
where $y_u$ is the up quark Yukawa coupling to the SM Higgs field $h$, and $\delta_{\rm br}$ is a phase angle of order unity.



Alternatively, we can consider a back reaction potential generated by an $SU(n_{HC})$ hidden color gauge interaction which confines at scales around the weak scale  \cite{relaxion, Antipin:2015jia}. For this, one can introduce the hidden colored matter superfields
\bea
L+L^c, \quad N+N^c
\eea
with the superpotential couplings
\bea
W_{\rm br} &=& \kappa_1 \frac{X_1^2}{M_*} LL^c + \kappa_uH_u LN^c + \kappa_d H_d L^c N,
\eea
where $L$ is an $SU(n_{HC})$-fundamental and $SU(2)_L$-doublet with the $U(1)_Y$ charge 1/2, $L^c$ is its conjugate representation,
$N$ is an $SU(n_{HC})$-fundamental but $SU(2)_L\times U(1)_Y$ singlet, and $N^c$ is its conjugate representation.
At scales  below $m_{\rm SUSY}$, all superpartners can be integrated out, leaving the following Yukawa interactions
between the relevant light degrees of freedom:
\bea
{\cal L}_{\rm br}\,=\, m_L e^{2i\phi_1/f_1}LL^c +\kappa_u\sin\beta hLN^c +\kappa_d \cos\beta h^\dagger L^cN + m_NNN^c,\eea
where  $L+L^c$ and $N+N^c$ denote the fermion components of the original superfields, $\tan\beta =\langle H_u\rangle/\langle H_d\rangle$, and
\bea
 m_L \,\sim\,  \kappa_1\frac{f_1^2}{M_*}\,\sim\, \kappa_1 m_{\rm SUSY}\eea
 is presumed to be lighter than $m_{\rm SUSY}$.
Note that a nonzero Dirac mass of $N+N^c$ is induced by radiative corrections below $m_{\rm SUSY}$, giving
\bea
m_N \,\sim\, \frac{1}{16\pi^2}\sin(2\beta)\kappa_u \kappa_dm_L^* e^{-2i\phi_1/f_1}\ln\left(\frac{m_{\rm SUSY}}{m_L}\right).
\eea
Now this effective theory at scales below $m_{\rm SUSY}$  corresponds to the non-QCD model proposed in \cite{relaxion, Antipin:2015jia}, yielding
a back reaction potential of the form
\bea
V_{\rm br}\,=\, m_1^2\,hh^\dagger \cos\left(2\frac{\phi}{f}+\delta_1\right) + m_2^4\cos\left(2\frac{\phi}{f}+\delta_2\right) ,\eea
where we have expressed the axion component $\phi_1$ in terms of the light relaxion field $\phi$, and
\bea
m_1^2 \,\sim\, \frac{\kappa_u\kappa_d\sin(2\beta)}{m_L}\Lambda_{\rm HC}^3, \quad m_2^4 \,\sim\, m_N\Lambda_{\rm HC}^3\eea
for the $SU(n_{HC})$ confinement scale $\Lambda_{\rm HC}$.
If $m_2^4 < m_1^2 v^2$ with $m_1^2 \lesssim {\cal O}(v^2)$, which can be achieved for $m_L< 4\pi v$ \cite{relaxion}, this back reaction potential can successfully stabilize the relaxion at a value giving $v=\langle h\rangle \ll m_{\rm SUSY}$.

\section{conclusion}
In this paper, we have addressed the problem of huge scale hierarchy in the relaxion mechanism, i.e. a relaxion excursion
$\Delta \phi \sim 2\pi f_{\rm eff}$ 
which is bigger than the period $2\pi f$ of the back reaction potential by many orders of magnitudes. We proposed a scheme
to yield an exponentially long relaxion direction within the compact field space of $N$ periodic axions with decay constants well below the Planck scale, giving
$f_{\rm eff}/f\sim e^{\xi N}$ with $\xi={\cal O}(1)$.
Although it relies on a specific form of the mass mixing between nearby axions, our scheme does not involve any fine tuning of continuous parameters, nor an unreasonably large discrete parameter. Furthermore, our scheme finds a natural UV completion in high scale or (mini) split SUSY scenario, in which all decay constants of the $N$ periodic axions are generated by SUSY breaking as  
$f_i \sim \sqrt{m_{\rm SUSY} M_*}$, where $m_{\rm SUSY}$ denotes the soft SUSY breaking scalar masses and $M_*$ is the fundamental scale such as the Planck scale or the GUT scale. In our model, the required relaxion potential and the relaxion-dependent Higgs boson mass are generated through a superpotential term providing a natural solution to the MSSM $\mu$-problem.

\section{acknowledgment}
We would like to thank Hyungjin Kim for helpful discussions. 
This work was supported by IBS under the project code, IBS-R018-D1.


\begin{thebibliography}{99}

\bibitem{relaxion} 
  P.~W.~Graham, D.~E.~Kaplan and S.~Rajendran,
  Phys.\ Rev.\ Lett.\  {\bf 115}, no. 22, 221801 (2015)
  [arXiv:1504.07551 [hep-ph]].
  
\bibitem{dvali} 
  G.~Dvali and A.~Vilenkin,
  Phys.\ Rev.\ D {\bf 70}, 063501 (2004)
  [hep-th/0304043].
  
\bibitem{Espinosa:2015eda} 
  J.~R.~Espinosa, C.~Grojean, G.~Panico, A.~Pomarol, O.~Pujolàs and G.~Servant,
  Phys.\ Rev.\ Lett.\  {\bf 115}, no. 25, 251803 (2015)
  [arXiv:1506.09217 [hep-ph]].
  
\bibitem{Hardy:2015laa} 
  E.~Hardy,
  JHEP {\bf 1511}, 077 (2015)
  [arXiv:1507.07525 [hep-ph]].
  
\bibitem{Patil:2015oxa} 
  S.~P.~Patil and P.~Schwaller,
  arXiv:1507.08649 [hep-ph].
  
  
\bibitem{Jaeckel:2015txa} 
  J.~Jaeckel, V.~M.~Mehta and L.~T.~Witkowski,
  arXiv:1508.03321 [hep-ph].
  
\bibitem{Gupta:2015uea} 
  R.~S.~Gupta, Z.~Komargodski, G.~Perez and L.~Ubaldi,
  arXiv:1509.00047 [hep-ph].
  
\bibitem{Batell:2015fma} 
  B.~Batell, G.~F.~Giudice and M.~McCullough,
  JHEP {\bf 1512}, 162 (2015)
  [arXiv:1509.00834 [hep-ph]].
  
\bibitem{Matsedonskyi:2015xta} 
  O.~Matsedonskyi,
  JHEP {\bf 1601}, 063 (2016)
  [arXiv:1509.03583 [hep-ph]].
  
\bibitem{Marzola:2015dia} 
  L.~Marzola and M.~Raidal,
  arXiv:1510.00710 [hep-ph].
 
  
  
\bibitem{Kim-Nilles-Peloso} 
  J.~E.~Kim, H.~P.~Nilles and M.~Peloso,
  JCAP {\bf 0501}, 005 (2005)
  [hep-ph/0409138].
  
\bibitem{Choi-Kim-Yun} 
  K.~Choi, H.~Kim and S.~Yun,
  Phys.\ Rev.\ D {\bf 90}, 023545 (2014)
  [arXiv:1404.6209 [hep-th]].
  
  
\bibitem{Kim-Nilles} 
  J.~E.~Kim and H.~P.~Nilles,
  Phys.\ Lett.\ B {\bf 138}, 150 (1984).

\bibitem{Murayama-Suzuki-Yanagida} 
  H.~Murayama, H.~Suzuki and T.~Yanagida,
  Phys.\ Lett.\ B {\bf 291}, 418 (1992).
  
\bibitem{Choi-Chun-Kim} 
  K.~Choi, E.~J.~Chun and J.~E.~Kim,
  Phys.\ Lett.\ B {\bf 403}, 209 (1997)
  [hep-ph/9608222].
  
\bibitem{Antipin:2015jia} 
  O.~Antipin and M.~Redi,
  JHEP {\bf 1512}, 031 (2015)
  [arXiv:1508.01112 [hep-ph]].
  
\bibitem{Intriligator:1995au} 
  K.~A.~Intriligator and N.~Seiberg,
  Nucl.\ Phys.\ Proc.\ Suppl.\  {\bf 45BC}, 1 (1996)
  [hep-th/9509066].
  
\bibitem{Grisaru:1996ve} 
  M.~T.~Grisaru, M.~Rocek and R.~von Unge,
  Phys.\ Lett.\ B {\bf 383}, 415 (1996)
  [hep-th/9605149].
  

\end{thebibliography}
\end{document}